\documentclass[12pt,preprint]{aastex}
\slugcomment{Submitted to ApJ}
\shorttitle{Quantitative Measurements of CME-driven Shocks}
\shortauthors{Ontiveros \& Vourlidas}

\begin{document}

\title{Quantitative Measurements of CME-driven Shocks from LASCO Observations}

\author{Veronica Ontiveros\altaffilmark{1}}
\affil{Instituto de Geofisica, Universidad Nacional Autonoma de
  Mexico, DF, 04510, MEXICO}
\email{vontiver@gmu.edu}
\and
\author{Angelos Vourlidas}
\affil{Code 7663, Naval Research Laboratory, Washington DC 20375, USA}
\email{vourlidas@nrl.navy.mil}

\altaffiltext{1}{present address: CEOSR, George Mason University, Fairfax VA, 22030, USA}

\begin{abstract}

In this paper, we demonstrate that CME-driven shocks can be
detected in white light coronagraph images and in which
properties such as the density compression ratio and
shock direction can be measured.  Also, their propagation direction can be
deduced via simple modeling. We focused on CMEs during the ascending
phase of solar cycle 23 when the large-scale morphology of the
corona was simple. We selected events which were good candidates to
drive a shock due to their high speeds (V$>$1500 km s$^{-1}$). The
final list includes 15 CMEs. For each event, we calibrated the LASCO
data, constructed excess mass images and searched for indications of
faint and relatively sharp fronts ahead of the bright CME front. We found such
signatures in 86\% (13/15) of the events and measured the
upstream/downstream densities to estimate the shock strength. Our
values are in agreement with theoretical expectations and show good
correlations with the CME kinetic energy and momentum. Finally, we
used a simple forward modeling technique to estimate the 3D shape
and orientation of the white light shock features. We found
excellent agreement with the observed density profiles and the
locations of the CME source regions. Our results strongly suggest that the observed brightness enhancements result from density enhancements due to a bow-shock structure driven by the CME.

\end{abstract}


\section{Introduction}

Coronal Mass Ejections (CMEs) are the largest transient expulsions of
coronal material in the heliosphere. These explosive events are
recorded by coronagraphs as brightness enhancements in white light
images because the ejected material scatters a large amount of
photospheric light. In image sequences, local brightness changes
provide most of the information on CME parameters, such as speed and
mass. LASCO \citep{bru95} observations have established that CME
speeds vary from a few hundred to more than 2500 km/s
\citep{yas04}. With this wide range, it is reasonable to expect that
CME speeds often exceed the local magnetosonic speed, and drive a
shock wave in the low corona \citep{hun87}.

There are two main observational results that provide support for the
existence of shocks in the low corona. Metric type-II radio bursts
provide indirect evidence of CME-driven shocks \citep[e.g.,][]{cli99},
but the scarcity of imaging radio observations precludes the reliable
identification of the driver. Observations of distant (from the CME)
streamer deflections \citep[e.g.,][]{gos74, mic84,she00} provide the
most reliable indication of a CME-driven wave pushing out the
streamers. However, there remains the question of whether this wave is
a shock wave, specially for the cases where the CME speed is not
excessively high. \cite{vou03} presented the first direct detection of
a CME-driven shock in white light images, combining two signatures:
(1) a sharp but faint brightness enhancement ahead of the CME and (2)
a streamer deflection well-connected to the expansion of the sharp
front. \cite{vou03} confirmed that the white light signature was a
shock wave using an MHD simulation based on the measured CME speed and
location.

Despite the large number of CME observations with LASCO, CME-driven
shocks in white light images remain difficult to detect. The
brightness enhancement due to the shock itself is faint and can be
easily lost in the background corona which changes from event to
event. Projection effects can also make it difficult to recognize and
separate the shock signatures from the rest of the CME because
deflected streamers, the shock and the CME material can all overlap
along a given line of sight (LOS). If the shock exists, however, it
will result in a density enhancement and, with proper analysis, should
be visible in the images.

We note that the earlier paper by \citet{vou03} discussed shock
signatures related to rather small and fast events (such as surges and
jets). Here we extend the detection of white light signatures to
standard CMEs by analyzing a set of fast CMEs during the rising phase
of solar cycle 23. These two papers suggest that white light shock
signatures must be a common feature in coronagraph images and the
previous scarcity of shock detections is mostly due to reduced
sensitivity, fields of view and temporal coverage of past
instruments. 

The data selection and methodology are described in
\S~\ref{selection}. The unique aspects of this work are the
quantitative density measurements that allow us to estimate the shock
strength, as presented in \S~3, and the analysis of the 3D morphology
and orientation of the white light shock using the Solar Corona
Raytrace (SCR), a software package that simulates the appearance of
various 3D geometries in white light coronagraph images
\citep{the06}. We compare the modeled images to density profiles
obtained from LASCO images in \S~\ref{shape} and found them in
excellent agreement. A summary and general discussion are presented in
\S~\ref{summary}.

\section{Event Selection and Identification of the White Light Shock \label{selection}}

To identify a sample of CME events with likely shock signatures, we
used two general criteria: (1) we searched for fast CMEs ($>$1500 km
s$^{-1}$) because they are more likely to drive a shock, and (2) we
considered only CMEs occurring at the ascending phase of the solar
cycle 23 (1997-1999), when the simple morphology of the background
white light corona offers a better chance to observe faint shock-like
structures with minimal confusion from overlapping structures along
the LOS.  Only 15 CMEs (out of about 2000) satisfied our selection
criteria and are shown in Table \ref{list}. The first and second
column are the number and the date of the event; the third column is
the time of first appearance in LASCO C2; the fourth, fifth and sixth
columns are the linear speed, angular width (AW) and the central
position angle (PA) respectively, as reported in the CME LASCO catalog
\citep{yas04}. The seventh column shows whether the CME is
associated with a decimetric type II radio burst \citep{gop05}.

To identify whether a shock signature exists in a given LASCO image,
we look for white light features that satisfy the following criteria:
(1) it must be a smooth, large scale front, (2) it must outline the
outermost envelope of the CME, and (3) it should be associated,
spatially and temporally, with streamer deflections. We choose these
criteria based on our expectations of how a CME-driven shock should
behave; namely, it should be ahead of the CME material (the driver),
it should expand away from the CME over large coronal volumes (but
avoiding coronal holes, for example), and it should affect streamers
when it impinges on them \citep{vou03}.

It turns out that such fronts exist in the majority of the images we
looked at but they are generally much fainter than most of the other
CME structures. They remain unnoticed in the running difference or
quick-look LASCO images normally published in the literature. These
fronts become visible only when the brightness scale of these images
is saturated to bring out the fainter structures.

\begin{figure}
\epsscale{0.8}
\plotone{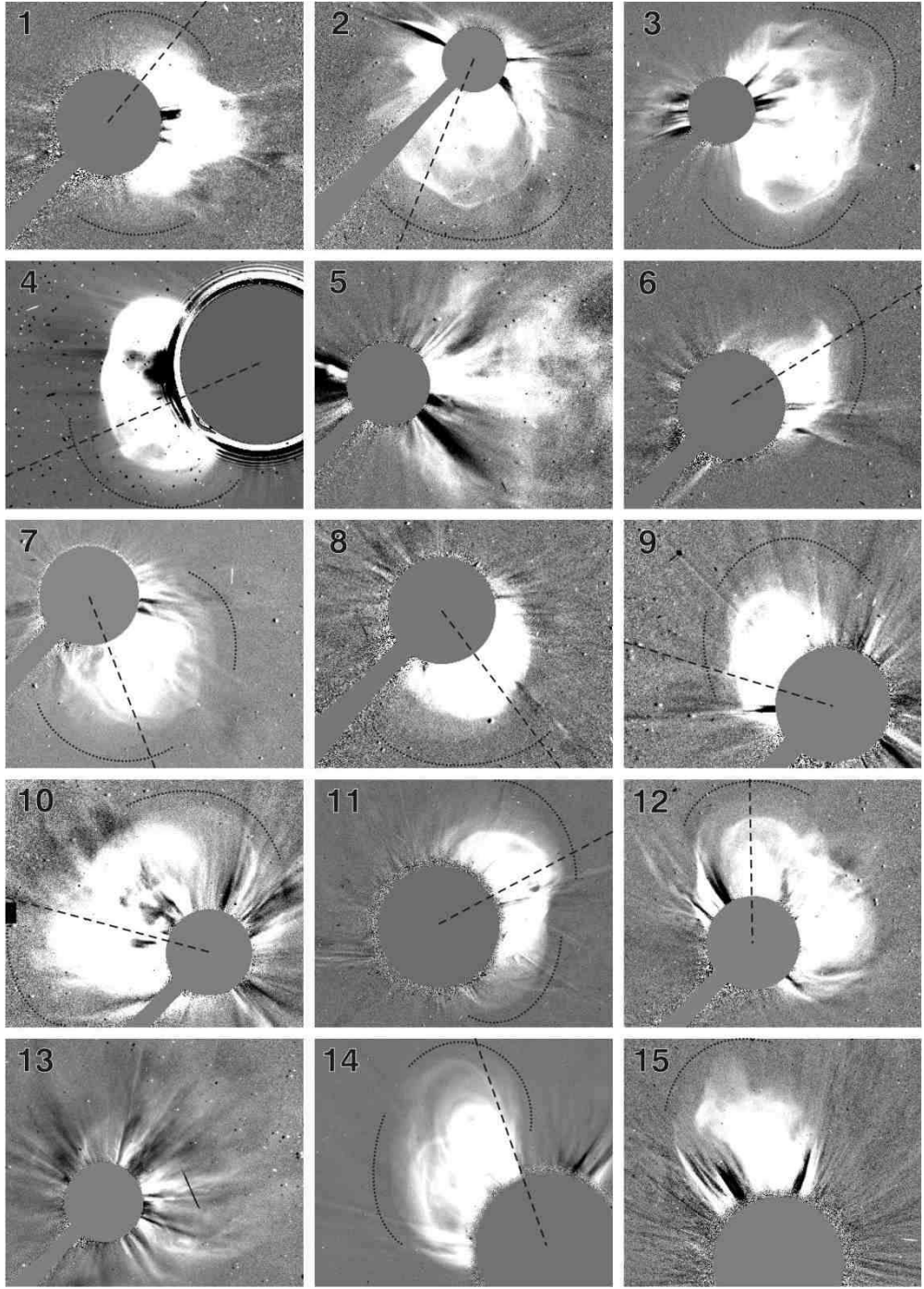}
\caption{The sample of the 15 fastest CMEs observed by the LASCO
  coronagraphs between 1997 and 1999. The image with the clearest
  shock signature is shown in each panel. The radial lines mark the
  position angles of the density profiles analyzed. The curved lines
  are visual guides for the location and extent of the faint shock
  structures. These features may be more visible in the online version
  of the figure. For events 5 and 13 it was not possible to determine
  a clear signature due to the disturbed background
  corona. \label{gallery}}
\end{figure}

We are able to identify and analyze these fronts because we use
calibrated LASCO images. The calibration brings out faint structures,
which may be missed in standard image processing, because it removes
vignetting and other instrumental effects.  We use excess
mass images from both LASCO C2 and C3. These are calibrated images
from which a pre-event has been subtracted thus removing the
background corona \citep{vou00}. A frame from each of our CMEs is
shown in figure \ref{gallery}. Most of the images have a curved line
to guide the reader's eye to the shock signature, while the dashed
line in each frame is pointing out the position angle corresponding to
the measurements that will be explained in section
\S~\ref{gammas}. Because of the large variation in the brightness of
the features, we have to apply different contrasts to bring out the
shock signature in each CME. In most images, the shock signature is
associated with diffuse emission on the periphery of the much brighter
CME material. The diffuse emission could arise from the coronal
material on the shock surface. For some events, like \#1, \#6, or
\#11, the faint emission encompasses the CME as one would expect for
an ideal case of a shock enveloping the driver. In other events, e.g
\#8, \#9 and \#12, the faint emission is only seen over a small range
of position angles. In all cases, the emission has a smooth front,
follows the general shape of the driver material (the bright CME
core), and is associated with a streamer bent from its pre-event
position. All these features are clear indications that we are dealing
with a wave. 

The reason why we are confident that what we observe is indeed a shock wave lies with the CME speed. All our events are either halo or partial halo CMEs, and their speeds are lower limits to the true speeds. Even these lower limit
speeds are sufficiently high to create a shock wave for typical values of plasma parameters in the low corona \citep{hun87}. Therefore, we will refer to the observed feature as shocks, from now on.

\section{Shock Measurements\label{res}}

\subsection{Shock Signature Evaluation\label{groups}}

We identified a shock signature in the LASCO images for 13 out of the
15 selected events. For events \#5 and \#13, it was not possible to
find a feature that satisfied even one of our three selection criteria
(\S~2). We believe that the lack of a smooth front and streamer
deflections may be due to the presence of a previous CME. In both
cases, the excess mass images showed clear evidence of a disturbed
corona (e.g., mass depletions, streamer displacements). A shock may
not form if the first event has altered the background magnetic field
considerably. Even if it forms, as the DM Type-II emissions suggest,
it is unlikely to develop a smooth, large front as it propagates
through such a disturbed medium. Similarly for streamer deflections,
many of the streamers could have already been deflected by the
previous CME at various angles from the sky plane and any new
deflection may not register as a smooth front in the images. Finally,
the strong intensity variations left on the image by the previous CME
may mask any faint fronts associated with the shocks from our CMEs. It
seems, therefore, that a relatively unperturbed corona facilitates the
detection of the faint CME-driven shocks. Nevertheless, once we establish which signatures are shock-related, we expect it will become easier to analyze events in more disturbed coronal conditions.
 
For events \#4, \#10 and \#11, we found a clear shock signature in the
LASCO C2 images, while for the rest of the events, the clearest
signatures were found in the LASCO C3 images. We found at least one
location with a clear white light shock signature for all halo CMEs
(10 events). This is expected if our shock interpretation is correct
since halos offer the best viewing of the CME flanks due to their
propagation along the LOS. We also note that our interpretation
implies that a major part of the halo CME extent is due to the shock
rather than actual ejected material and analysis of CME widths need to
take this fact into consideration.

\subsection{From Mass Profiles to Density Ratios\label{dens}}

We can use the calibrated LASCO images not only to identify the faint
shock fronts but also to derive some estimates of the density profile
across these fronts. Each pixel in the mass LASCO images gives the
total mass or, equivalently, the total number of electrons along the
LOS. This excess electron column density (e/cm$^{2}$) can be converted
to electron volume density (e/cm$^{3}$), if the depth of the structure
along the LOS is known. This quantity is unknown and it cannot be
reliably estimated without some knowledge of the 3D morphology of the
structure. We address the 3D aspect in \S~4 but for the analysis we
assume a nominal depth of 1 R$_\sun$ for all the events because it is
a convenient scale and likely a good upper limit given the slope
($\sim 0.3 R_\sun$) of the brightness profiles (e.g.,
figure~\ref{ev1_2}).

We then derive the total electron column density along the profile, by
integrating the density of the background equatorial corona from the
Saito, Poland, Munro (SPM) model \citep{sai77}. Again, the actual
value of the background density for each event is not known and we
have to resort to a density model as is often the case in analysis of
coronal observations. Here we assume that the same equatorial SPM
model applies to all events for two reasons: (1) our sample covers
only a small phase of the solar cycle when the average density of the
background corona does not vary significantly, and (2) the density
enhancement at the shock front must come from streamer material since
shocks do not propagate, nor pile up material over coronal holes.

\subsection{Estimation of the Shock Strength\label{gammas}}

For each event we chose the image with the visually clearest shock
signature. In that image, we obtain several profiles at different PAs
along the shock front. Our method averages the emission along a narrow
range of PAs ($\sim$5$\degr$) to improve the signal-to-noise
ratio. The radial extent of each profile allows us to obtain the
upstream and downstream brightness at different angles of the shock
\citep[see also][]{vou03} and convert them to densities as mentioned
in \S~3.2. We classify the 15 events in four groups based on the
appearance of the shock signatures in the images and the evidence of a
jump in the density profiles.  Group 1 includes those events which
have a clear shock signature in the image and a steep jump in the
brightness profiles at the location of the shock front (6 events).
Group 2 includes those events which show a clear shock signature in
the image but the density jump at the shock front is barely detectable
above the noise (5 events).  Group 3 includes those events which have
shock signatures in the image but the density profiles are too noisy
to identify the jump at the shock front; there were 2 events in Group
3.  Finally, Group 4 are the two events (\#5 and \#13) without any
shock signatures in any LASCO image.

The shock fronts are more visible on the images rather than in the
density profiles because of our eyes' spatial averaging ability. We
believe that the density profiles can be improved by averaging over a
larger angular width. However, this averaging tends to smooth the
profiles and reduce the density jump. Until we find a better averaging
method we adopted the $5 \degr$ width in the current work.

For this reason, we concentrate on the profiles with the sharpest
density jump. We use the density jump as a proxy to the shock
strength. We define $\Gamma_{CR}$ as the compression ratio of the
total to background volume densities, $\Gamma_{CR} = 1
+\frac{\rho}{\rho_0}, $ where $\rho$ is the excess density due to the
shock, and $\rho_0$ the upstream density obtained from the SPM model.

\begin{figure}
\plotone{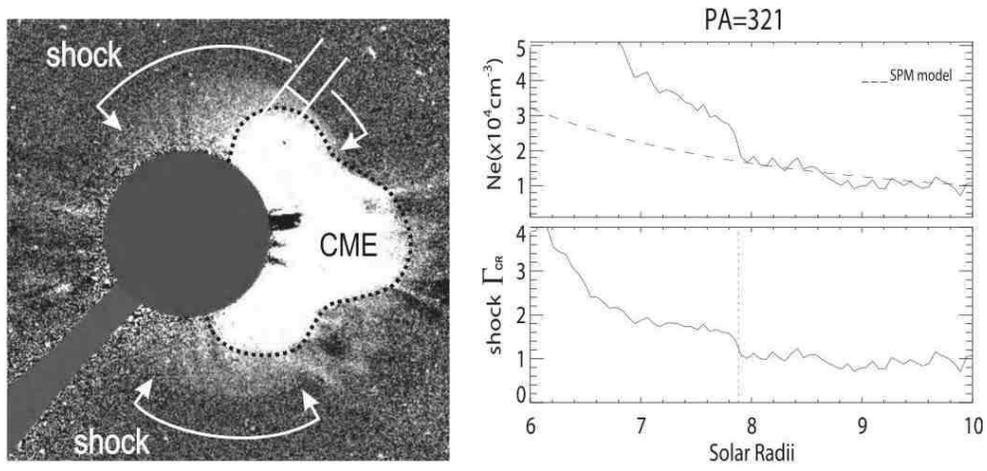}
\caption{\textit{Left panel:\/} Selected image for the November 6, 1997
  CME. A clear shock signature can be seen at the flanks of the
  CME. The parallel lines over the shock front show the profile with the strongest shock signature. \textit{Right panel:\/} The top plot shows the estimated
  up-stream and down-stream density profile at PA = 312 (solid) and the
  background corona density from the SPM model (dashed). The bottom
  plot shows the density ratio, $\Gamma_{CR}=1.6$, at $7.9$
  $R_\sun$ which we use as a proxy to the shock strength.}\label{ev1_2}
\end{figure}

Figure \ref{ev1_2} shows event \#1 as an example. The two parallel
lines mark the position angles we average over to obtain the
brightness profile with the sharpest density jump, and hence the
strongest shock signature. The jump is located at 7.9 R$_\sun$ at
PA$=$321$\degr$. The observed density profiles and the ratio between
excess and background densities are plotted on the right panels of
figure \ref{ev1_2}. In this case, we obtain a $\Gamma_{CR}$ of 1.6 at
the location of the shock. We repeat the same analysis for the
remaining events in our sample. We also obtain the CME mass, momentum
and kinetic energy from the same images following the method described
by \cite{vou00} and using the speed measurements from the LASCO CME
catalog. These measurements allow us to get global CME parameters to
compare with the local shock strength, which are discussed next.

\subsection{Statistics} \label{statistics}

Table \ref{results} presents our measurements for the 13
events. Columns 1-2 correspond to the event number (\ref{list}) and
its group (~\S\ref{groups}). Columns 3 to 6 are the time of LASCO
observation, the heliocentric distance to the shock signature, the PA
of the profile and the estimated density jump, $\Gamma_{CR}$, of the
shock for that profile. Columns 7 to 9 are the CME mass, kinetic
energy and momentum respectively, obtained from calibrated images. Now
we can assess the validity of our main assumption; namely, whether the
faint structures seen ahead of the main CME ejecta could indeed be the
white light counterpart of the CME-driven shock. If this is true, we
expect a correlation between the magnitude of the density jump (or
$\Gamma_{CR}$) and the CME dynamical parameters, such as the CME
kinetic energy.

\begin{figure}
\epsscale{0.4}
\plotone{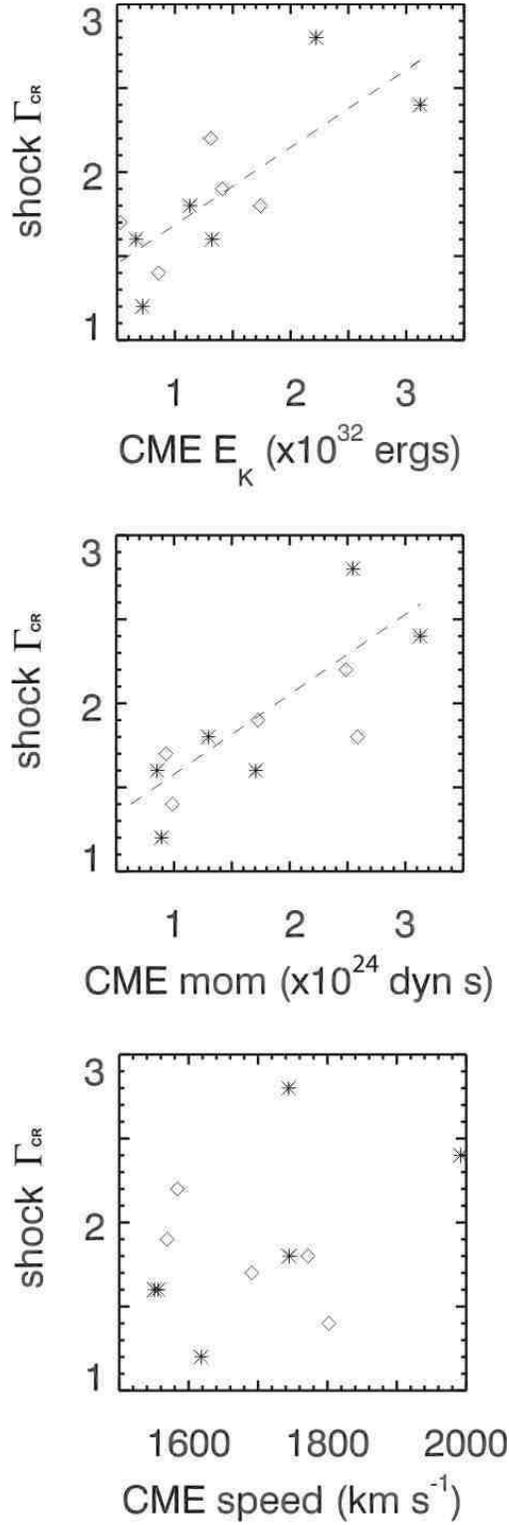}
\caption{Trends and correlations between estimated shock $\Gamma_{CR}$ and
  select CME parameters. The top and middle panels show the correlations between
  the shock $\Gamma_{CR}$ and the CME kinetic energy (cc=0.77) and momentum
  (cc=0.80), respectively. The bottom panel shows no clear correlation
  between the shock $\Gamma_{CR}$ and the CME speed (group 1: stars, group
  2: diamonds). \label{trends}}
\end{figure}

The plots in figure \ref{trends} show the trends and correlations
obtained between $\Gamma_{CR}$ and some CME parameters for the best
events only (groups 1 and 2). We find important correlations between
the shock strength and the CME momentum (cc=0.80), and kinetic energy
(cc=0.77). Furthermore, the largest $\Gamma_{CR}$ are associated with
the sharpest shock signatures (group 1, see \S~\ref{groups}) and the
highest kinetic energies and momenta. These results suggest that our
$\Gamma_{CR}$ parameter is associated with the CME dynamics as
expected from a shock-produced density jump.

Perhaps surprisingly, there is no obvious correlation between the
strength and the linear speed of the CMEs. The quoted speeds are taken
from the LASCO catalog and therefore correspond to the speed at the
position angle of the fastest moving feature in the LASCO images. Our
profiles were taken at different position angles since the shock front
is more easily discernible at some distance away from the CME
front. Also, the speeds are derived from linear fits to height-time
measurements extending over the full range of the LASCO field of view
(2-30 R$_\sun$) and correspond to the average CME speed over the field
of view. Our density jump is derived from a single snapshot of the CME
at a single heliocentric distance. A plot between $\Gamma_{CR}$ and
the CME speed at the same PA and distance might provide a better
correlation. We tried to make these speed measurements. However, the
large CME speed and synoptic LASCO cadences did not allow us to obtain
a sufficient number of data points to derive reliable speed
estimations for any of our events.

\begin{figure}
\plotone{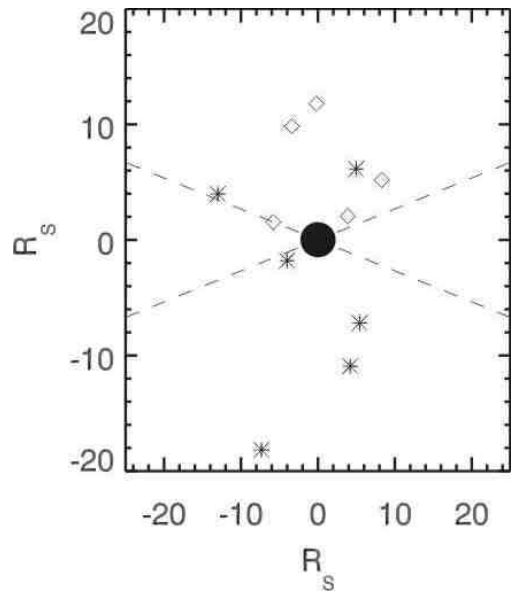}
\caption{Locations on the plane of the sky of the clearest shock
  signatures in our event list. The dashed lines separate favorable
  and non-favorable position angles for a shock observation. All
  events showing a clear shock signature in the image and/or in the
  data analysis (stars and diamonds labels), are found below or above
  15 degrees from the solar equator.\label{polar}}
\end{figure}

Figure~\ref{polar} is related to the visibility of the shock signature
in the LASCO images and shows that the clearest shock signatures were
found above or below $15\degr$ with respect to the solar equator. This
holds even for the halo CMEs where there the shock is visible over
more position angles. Considering the phase of the solar cycle, these
results show that locations away from the streamers are favorable
angles for shock signature observations on white light images. This
result should be kept in mind when searching for shock signatures in
coronagraph images. The complexity of the background corona masks the
faint shock emissions during solar maximum while there are few
sufficiently fast events to drive a shock during solar minimum.

\section{Estimating the Shock Geometry with a Forward-Modeling
  Technique\label{shape}}

The results in the previous sections provide ample support for a shock
interpretation of the faint emission ahead of the CME front. In addition, we have devised a practical way to model the faint emission with a prescribed shape using
forward modeling techniques. The advantages of this approach are the
speed and simplicity of the software, and the resulting information on
the 3D morphology and direction of the shock. The disadvantage is that
we have no way of calculating a goodness of fit for the model other
than a visual judgment on whether the envelope of the model fits the
observed emission envelope. For the sake of brevity, we use the term
'fitting' from now on to describe the 'fitting-by-eye' we actually
employed.

For simplicity, we assume that the CME-driven shock has a 3-D bow
shock morphology, as expected from a body moving in uniform magnetized
flow (e.g., Earth's magnetosphere). We first need a geometric
description for such a model. We found one in \citet{smith03} which is
used to describe the shocks around Herbig-Haro objects. It is a a
surface of revolution, in cylindrical coordinates and is described by
the form (Eq. (1) in \citet{smith03})
\begin{equation}
\frac{z}{d} = \frac{1}{s} (\frac{R}{d})^s
\end{equation}
where s controls the opening angle of the bow, d is a scale length
(\textsl{semilatus rectum\/}) and $R$ is the heliocentric distance of
the nose of the shock. The variables $d$ and $s$ control the shape of
the bow shock and are the most important variables for visually
matching the shock shape to the observations. To adopt this model for
coronagraphic observations, we add a narrow shell of constant density,
$N_{e0}$, and width, $\Delta$. In this way, we can calculate the
brightness of the model using the Thomson scattering equations and the
spacecraft geometry and analyze the model images exactly as we do the
observations. We justify this shell as the plasma enhancement around
the shock surface at a given moment. While plasma pileup at CME fronts
is still an open question \citep{howard05}, it is expected that the
shock will cause local density enhancements as it propagates through
the corona \cite{vou03}.

We use the SCR software package to create a simulated coronagraph
image from the model. SCR is a numerical implementation of Thomson
scattering that renders a total (or polarized) brightness 2D image as
seen by a coronagraph (e.g. LASCO C2 or C3) given a 3D density
structure distribution \citep{the06} and is available in
Solarsoft. For all events, we assumed a constant thickness of $\Delta
= 0.3 R_\sun$ which is comparable to the width of the brightness jump
in the images. Because the height of the shock varied for each event,
we set the density, $N_{e0}$ within the thin shell to the estimated
density just ahead of the shock front (\S~3.2). Once the width and
density of the model shock were set, we tried to match it to the LASCO
images by varying its orientation in space and the geometric
parameters of the parabola ($d$ and $s$). When we were satisfied that
the simulated image fit visually the observed envelope of the shock,
we integrated along the 3D shape, using the LASCO viewing geometry,
and obtained a simulated brightness image of the model shock.

\begin{figure}
\plotone{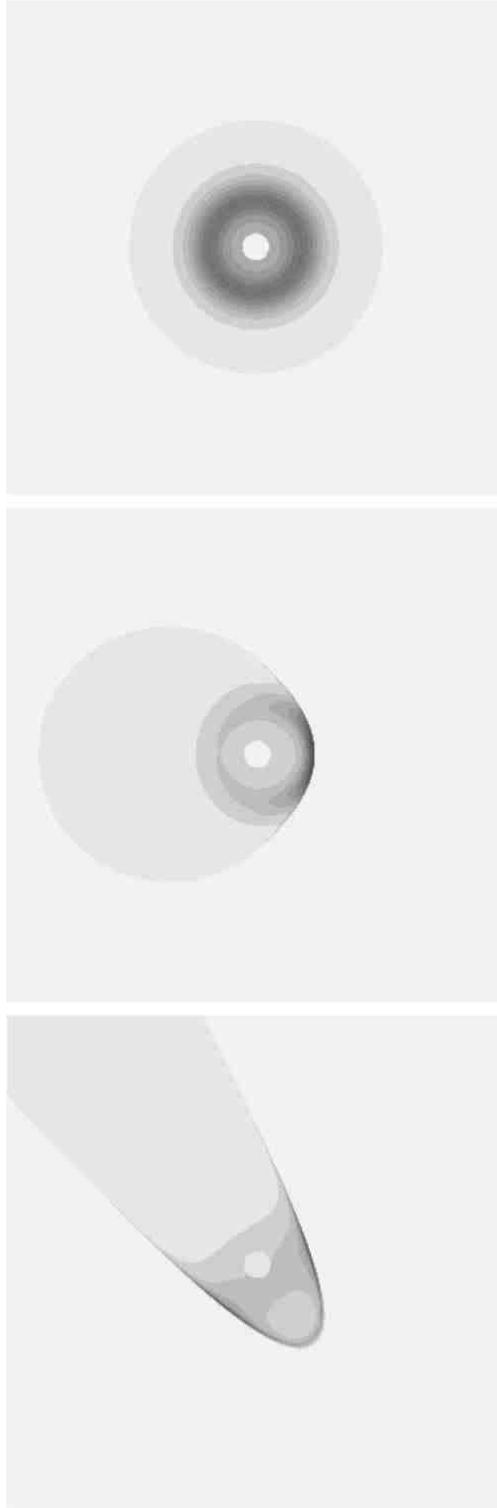}
\epsscale{0.4}
\caption{Simulated white light images for a bow-shock model at 8
  R$_\sun$ observed through different lines of sight, from top to
  bottom:(1) Along the Sun-Earth line, (2) 10$\degr$ west and (3)
  45$\degr$ west, 45$\degr$ south. The hole in the center of the
  images shows the size of the solar disk for scale. The intensity
  gradient represents the white light brightness of the model as
  viewed by LASCO C3.\label{los}}
\end{figure}

Figure \ref{los} shows simulated white light images of our bow-shock
model viewed from different lines of sight: (1) Along the Sun-Earth
line, (2) $10\degr$ west and (3) 45$\degr$ west, 45$\degr$ south. The
location of the bow shock nose is at 8 $R_\sun$. The images show the
full model for completeness but we have restricted our integrations to
a volume of $30 R_\sun^3$, so the actual model is truncated and the
long thin extensions in Figure~\ref{los} do not appear in our images.

\begin{figure}
\epsscale{0.7}
\plotone{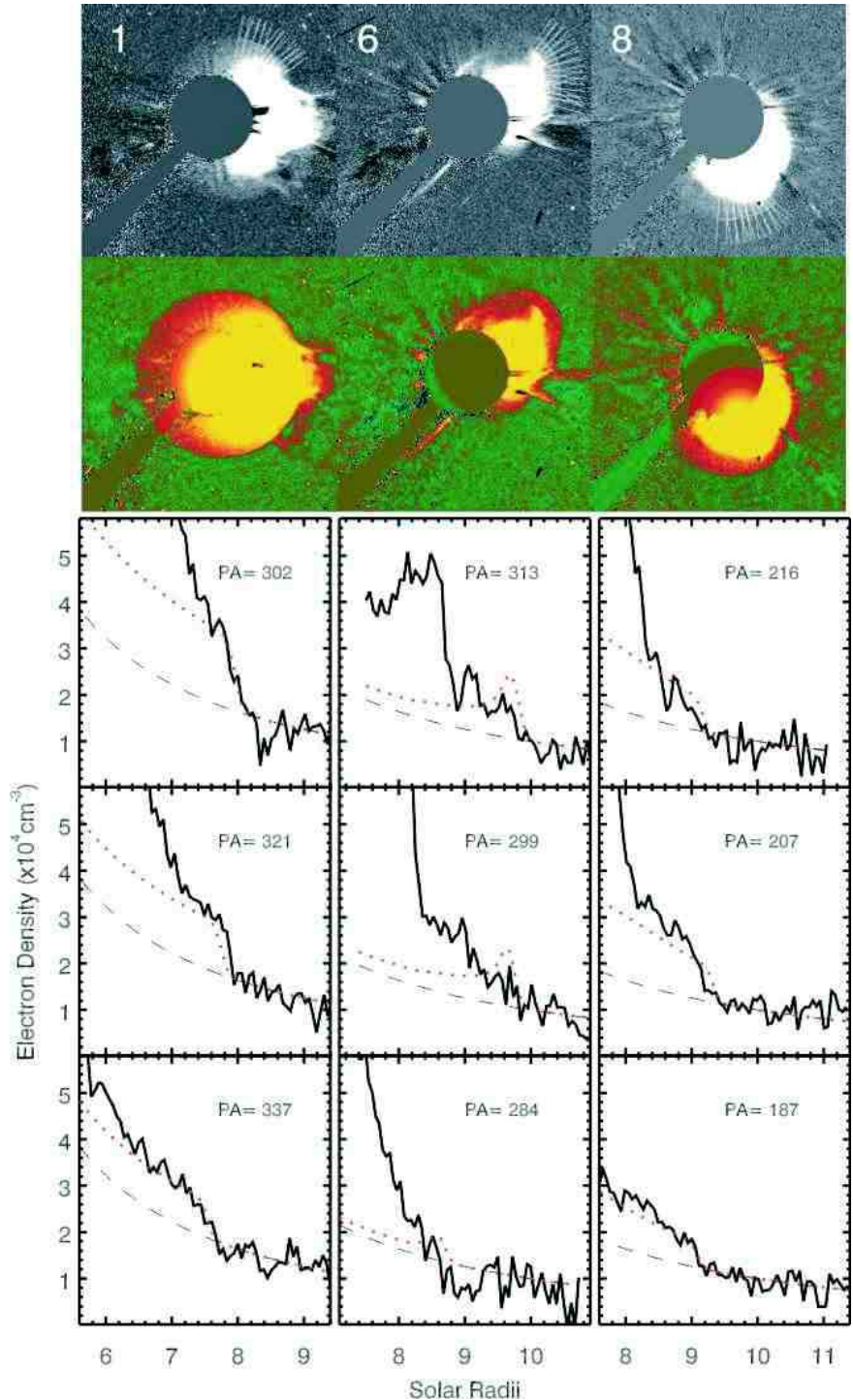}
\caption{Comparing observed and modeled density profiles for events \#1,\#
  6 and \#8. From top to bottom for each column:(1) the actual LASCO
  image and the selected angles for obtaining the density profiles;
  (2) A synthetic coronagraph image of the shock model overplotted on the LASCO
  image; (3-5) comparison of density profiles at different PAs (solid
  line: LASCO density profile; dotted line: model shock density profile; dashed
  line: background coronal density from the SPM model).\label{ray}}
\end{figure}

To check whether the LASCO density profiles are consistent with a
bow-shock geometry, we fit an SCR model to each event and obtained
simulated density profiles over several PAs, using the same method as
described in \S~\ref{dens} for the LASCO observations. This procedure
is currently done by hand and the LOS integration is time-consuming in
our hardware. We are working on improving it but so far we were able
to perform detailed comparisons for only three events in our
sample. We chose events \#1 (November 6, 1997), \#6 (June 4, 1998) and
\#8 (November 26, 1998), which have some of the clearest shock
signatures. Figure \ref{ray} shows the results. Each plot shows the
comparison between the SCR profiles to the LASCO ones for different
PAs at the shock signature.

Figure \ref{ray} shows that the observed density profiles are
consistent with a bow shock geometry for at least 30$\degr$ along the
shock signature. The density fits are surprisingly good given the
simplicity of our model. Note that we did not attempt to fit either
the density jump or the shock LOS extent. They were kept constant for
each event. This result offers a strong indication that the overall
envelope of enhanced emission around the CME must come from a simple
structure (e.g., a bow shock) probably reflecting the simplicity of
the minimum corona. In other words, the shock structure, and probably
its visibility, may depend on the overall configuration of the large
scale corona. It will be interesting to repeat our analysis for CMEs
during the solar maximum.

Another consistency check comes from comparing the orientation of the
bow shock shell in 3D space (which we get from the SCR modeling) to
the expected orientation of the CME. For this, we make the usual
assumption that the core of the CME (the driver for us) propagates
radially outwards from the nearest possible source region (e.g., a
flaring active region). The source regions for our three events are as follows:
\begin{itemize}
\item Event \#1 is associated with the flare
observed at 11:49 at S18W63. We took this location as the source
region of the CME. Since this is a front side event, it is relatively
easy to determine that there is no other active region with a better
association.
\item The LASCO movies suggest that event \#6 is likely associated with a
filament eruption on the far side of the Sun. The filament was seen
for several days in H$\alpha$ images as it crossed over the western
limb. Extrapolating from its known position on May 29th we estimate
that the center of the filament would be at N43W107 on the day of the
eruption.
\item Event \#8 is also a backside CME resulting in an indirect source
region identification. We examined the EIT and LASCO movies for a few
days before and during the eruption. The low corona signatures of the
eruption suggest that active region 8384 is the most likely source
and should be located around S26W134 at the time of the eruption.
\end{itemize}

We then calculated the heliographic coordinates of the nose of our
modeled bow shock for each of the three events. The results for the
three events are shown in Table~\ref{location}. Again, we did not
attempt to take into account the location of the source region when we
fit the geometric model. Only during the writing of this paper we
calculated the final position of the shock nose and
compared it with the possible source regions. We were surprised to
find that the direction of the modeled shock is within $30\degr$ of the
expected CME nose, assuming radial propagation from the source
region. The discrepancy could be simply due to non-radial expansion of
the CME or uncertainty in the source region since two of the events
were backside CMEs. Given these restrictions, the results in
Table~\ref{location} are very encouraging because they suggest that
our forward modeling approach can provide useful information on the 3D
morphology and orientation of the shock using a single viewpoint and
modest hardware and software resources. We plan to investigate the
sensitivity of the derived shock orientation to different model fits
and apply it to more events in a future paper.

\section{Summary and Discussion \label{summary}}

In this paper, we demonstrate that the CME-driven shock is indeed
visible in coronagraph images. It can be seen as the faint large scale
emission ahead and around the bright CME material. To establish this
interpretation, we started by selecting all fast CMEs ($>$1500 km s$^{-1}$)
observed by LASCO between 1997-1999 (15 events). We found the
following:
\begin{itemize}
\item Ten of our events are associated with a decametric type II radio
  burst, suggesting the existence of a shock wave in the outer
  corona. The remaining five events are backside CMEs where
  the detection of radio burst is not always possible and the
  existence of a shock cannot be ruled out. In other words, the
  existence of a shock at the heights of the LASCO observations (2-30
  Rs) is supported by other observations for all events in our sample.
\item 86\% of these events exhibited a relatively sharp but faint brightness
  enhancement ahead or at the flanks of the CME over a large area,
  which we interpret as the white light counterpart of the CME-driven shock.
\item All halo CMEs (10 events) have at least one location with such a
  shock signature. This is consistent with a shock draping all around
  the CME driver.
\item The clearest white light signatures were found $15^\circ$ above
  or below the solar equator, irrespective of heliocentric
  distance. It is possible that the morphology and complexity of the
  corona along the LOS plays a role in identifying the shock in white
  light images.  The two events with no white light shock signatures
  were also the fastest, came in the wake of a previous large-scale
  CME. As we discussed in \S~3.1, the disturbed background corona may
  be responsible for the lack of shock signatures. It is also likely
  that any shock signatures may have been missed because of the low
  observational cadence and high speed of these events.
\item We found only a weak dependence between the shock strength
  ($\Gamma_{CR}$) and the CME speed. There may be several reasons for this
  discrepancy: (1) the speeds are more sensitive to projection effects, (2) the strength and speeds are measured at different PAs
  and/or (3) the speeds correspond to the average CME speed in the LASCO
  field of view while the strengths are measured in a single image.
\item We found stronger correlations between the density jump and the
  CME kinetic energy (cc=0.77) and between density jump and the momentum
  (cc=0.80). This is a very encouraging result because it shows that
  our density jump is closely related to the CME dynamics and hence
  more likely to correspond to a true shock jump.
\item We are able to account for the smooth observed jumps in the
  brightness profiles (and the derived density profiles) as compared to the
  step-like jumps observed in-situ. We found that they can be
  reproduced by a line-of-sight integration through a thin ($\sim 0.3
  R_\sun$) shell of material. This material is presumably the locally
  enhanced corona, which has become compressed due to the passage of the shock.
\end{itemize}

The high CME speeds, the sharpness of the features and the brightness
jumps are all strong indicators that our interpretation of these
features as the white light counterpart of CME-driven shocks is
correct. The strong correlations of the density jump to the CME
kinetic energy and momentum provide additional support. Based on the
information presented here, it should be a
simple matter to identify such features in all events where a shock is
expected. We have found many more examples in a quick
survey of LASCO images throughout the mission. It is still difficult,
however, to extract quantitative measurements from all of these shocks
due to the lower signal-to-noise ratio of the individual density
profiles compared to the images. We are looking for ways to
average across the shock front without introducing unnecessary
smoothing to it.

We have examined whether the observed shock shapes are consistent with
expected 3D shock geometries. We used a standard bow shock geometric
model, adapted from astrophysical shocks, and a forward modeling
software package from the SECCHI solarsoft collection to test a quick
method of estimating the shock size and orientation for
coronagraphs. We found that a bow-shock geometry is indeed a good fit
to the observed LASCO morphology and it readily explains the observed
density profiles. The simulated profiles can match the observed
profiles over several position angles, and even at large heliocentric
distances. We also found that our modeled 3D shock direction is in
fairly good agreement with the expected direction of the CME assuming
radial propagation from the source region.

These results suggest that we can not only estimate the 3-dimensional
shape and direction of the CME-driven shock but we can also use the
model fits to separate the brightness enhancement of the shock from
that of the driving material and thus obtain more accurate
measurements of the CME and shock characteristics. One such quantity
is the shock kinetic energy which plays an important role in
understanding and modeling the production of solar energetic particles
from shocks. We will pursue these ideas further in a future paper.

\acknowledgments We thank R. A. Howard for his editorial help and
valuable comments, which have improved the text considerably;
A. F. Thernisien for his generous help with the SCR software, and the
anonymous referee for his/her careful reading and suggestions. This work
is funded by the LWS TR\&T grant NNH06AD851. The CME catalog is
generated and maintained by NASA and The Catholic University of
America in cooperation with the Naval Research Laboratory. SOHO is a
project of international cooperation between ESA and NASA.

\clearpage

\begin{deluxetable}{lcccccc}
\tabletypesize{\scriptsize}

\tablecaption{All high speed CMEs (V$>$1500 km s $^{-1}$) between 1997-1999\label{list}}
\tablewidth{0pt}
\tablehead{
\colhead{event} & \colhead{CME date} &\colhead{1st appearance} & \colhead{linear speed} & \colhead{AW} & \colhead{PA} &\colhead{type II}\\
& \colhead{(yymmdd)} &\colhead{(C2 UT)} & \colhead{(km s$^{-1}$)} & \colhead{(deg)} & \colhead{(deg)}&\colhead{(Dm)}
}
\startdata
1 & 971106 & 12:10:00 & 1556 & 360 & 262 & yes\\
2 & 980331 & 6:12:00 & 1992 & 360 & 177 & no\\
3 & 980420 & 10:07:00 & 1863 & 165 & 278 & yes\\
4 & 980423 & 5:27:00 & 1618 & 360 & 116 & yes\\
5 & 980509 & 3:35:58 & 2331 & 178 & 262 & yes\\
6 & 980604 & 2:04:00 & 1802 & 360 & 314  & no\\
7 & 981124 & 2:30:00 & 1798 & 360 & 226  & no\\
8 & 981126 & 6:18:05 & 1505 & 360 & 198 &no\\
9 & 981218 & 18:21:00 & 1749 & 360 & 36 & yes\\
10 & 990503 & 6:06:00 & 1584 & 360 & 88 &yes\\
11 & 990527 & 11:06:00 & 1691 & 360 & 341 &yes\\
12 & 990601 & 19:37:00 & 1772 & 360 & 359 &yes\\
13 & 990604 & 7:26:54 & 2230 & 150 & 289 &yes\\
14 & 990611 & 11:26:00 & 1569 & 181 & 38 & yes\\
15 & 990911 & 21:54:00 & 1680 & 120 & 13 & no\\
\enddata

\end{deluxetable}

\clearpage

\begin{deluxetable}{llccccccc}
\tabletypesize{\scriptsize}

\tablecaption{Results\label{results}}
\tablewidth{0pt}
\tablehead{
\colhead{event} & \colhead{group} &\colhead{shock time} & \colhead{shock height} & \colhead{shock PA} & \colhead{$\Gamma_{CR}$} & \colhead{mass} & \colhead{kinetic energy} & \colhead{momentum}\\
& &\colhead{(UT)} & \colhead{(R$_\sun$)} & \colhead{(deg)} & &\colhead{($\times$10$^{15}$g)}& \colhead{($\times$10$^{31}$ erg)} & \colhead{($\times$10$^{24}$ dyn s)}}
\startdata
1 & 1 & 12:41:05 & 7.9 & 321 & 1.6 & 5.48 & 6.63 & 0.85 \\
2 & 1 & 7:29:37 & 19.6 & 158 & 2.4 & 15.74 & 31.23 & 3.13 \\
3 & 3 & 12:42:05 & 23.7 & \nodata & \nodata & 23.52 & 40.82 & 4.38 \\
4 & 1 & 5:55:22 & 4.4 & 114 & 1.2 & 5.51 & 7.21 & 0.89 \\
5 & 4 & \nodata & \nodata & \nodata & \nodata & \nodata & \nodata & \nodata \\
6 & 2 & 3:41:14 & 9.8 & 302 & 1.4 & 5.3 & 8.6 & 0.95\\
7 & 1 & 4:42:05 & 11.7 & 201 & 2.8 & 14.62 & 22.23 & 2.55\\
8 & 1 & 6:18:05 & 9 & 217 & 1.6 & 10.98 & 13.21 & 1.70\\
9 & 1 & 19:41:42 & 13.6 & 73 & 1.8 & 7.43 & 11.32 & 1.30\\
10 & 2 & 8:18:05 & 6 & 75 & 2.2 & 10.44 & 13.1 & 1.65\\
11 & 2 & 13:38:17 & 4.4 & 298 & 1.7 & 3.7 & 5.28 & 0.63\\
12 & 2 & 21:18:07 & 11.8 & 1 & 1.8 & 11.09 & 17.41 & 1.96\\
13 & 4 & \nodata & \nodata & \nodata & \nodata & \nodata & \nodata & \nodata \\
14 & 2 & 14:18:05 & 10.4 & 19 & 1.9 & 11.42 & 14.06 & 1.79\\
15 & 3 & 23:42:05 & 17.9 & \nodata & \nodata & 2.6 & 3.67 & 0.44\\
\enddata

\end{deluxetable}

\clearpage
\begin{deluxetable}{lrr}
\tabletypesize{\scriptsize}

\tablecaption{Comparison of Modeled Shock Orientation and CME Source Regions\label{location}}
\tablewidth{0pt}
\tablehead{
\colhead{event} & \colhead{shock nose} &\colhead{source region}\\
}
\startdata
971106 & S13W56 & S18W63 \\
980604 & N47W138 & N43W107 \\
981126 & S38W108 &S26W134 \\
\enddata
\end{deluxetable}
\clearpage
\clearpage
\clearpage

\clearpage

\clearpage
\clearpage

\end{document}